\documentclass[lettersize,journal]{IEEEtran}
\usepackage{amsmath,amsfonts}
\usepackage{algorithmic}
\usepackage{algorithm}
\usepackage{array}
\usepackage{textcomp}
\usepackage{stfloats}
\usepackage{url}
\usepackage{verbatim}
\usepackage{graphicx}
\usepackage{cite}
\usepackage{array}
\usepackage{mdwmath}
\usepackage{mdwtab}
\usepackage{eqparbox}
\usepackage{url}
\usepackage{amssymb}
\usepackage{amsmath}
\usepackage{booktabs}
\usepackage{subfigure}

\begin{document}

\title{Capacity Analysis of Holographic MIMO Channels with Practical Constraints}

\author{Yuan Zhang,
Jianhua Zhang,~\IEEEmembership{Senior Member,~IEEE,}
Yuxiang Zhang,~\IEEEmembership{Member,~IEEE,}
Yuan Yao,~\IEEEmembership{Senior Member,~IEEE,}
and Guangyi Liu,~\IEEEmembership{Member,~IEEE}

\thanks{Manuscript received xx; revised xx; accepted xx. Date of publication xx; date of current version
xx. This work was supported in part by the National Science Fund for Distinguished Young Scholars under Grant 61925102; in part by the National Natural Science Foundation of China under Grant 92167202; in part by the National Natural Science Foundation of China under Grant 62101069; in part the National Key Research and Development Program of China under Grant 2020YFB1805002; and in part by the BUPT-CMCC Joint Innovation
Center. The associate editor coordinating the review of this article and
approving it for publication was xx. (Corresponding author: Jianhua Zhang.)

Yuan Zhang, Jianhua Zhang, Yuxiang Zhang, and Yuan Yao
are with Beijing University of Posts and Telecommunications, Beijing 100876,
China (e-mail: yuanzhang, jhzhang, zhangyx, yaoy@bupt.edu.cn).

Guangyi Liu is with China Mobile Research Institute, Beijing 100053, China (liuguangyi@chinamobile.com).}
}

\markboth{}%
{Shell \MakeLowercase{\textit{et al.}}: A Sample Article Using IEEEtran.cls for IEEE Journals}


\maketitle

\begin{abstract}
Holographic Multiple-Input and Multiple-Output (MIMO) is envisioned as a promising technology to realize unprecedented spectral efficiency by integrating a large number of antennas into a compact space. Most research on holographic MIMO is based on isotropic scattering environments, and the antenna gain is assumed to be unlimited by deployment space. However, the channel might not satisfy isotropic scattering because of generalized angle distributions, and the antenna gain is limited by the array aperture in reality. In this letter, we aim to analyze the holographic MIMO channel capacity under practical angle distribution and array aperture constraints. First, we calculate the spectral density for generalized angle distributions by introducing a wavenumber domain-based method. And then, the capacity under generalized angle distributions is analyzed and two different aperture schemes are considered. Finally, numerical results show that the capacity is obviously affected by angle distribution at high signal-to-noise ratio (SNR) but hardly affected at low SNR, and the capacity will not increase infinitely with antenna density due to the array aperture constraint.
\end{abstract}


\section{Introduction}
With the increasing demand for data transmission, higher spectral efficiency technologies attract great interest. Holographic Multiple-Input and Multiple-Output (MIMO) is expected to realize incredible spectral efficiency based on the excellent performance of massive MIMO technology \cite{marzetta2010noncooperative}. Compared with traditional massive MIMO systems, holographic MIMO systems are equipped with denser (possibly infinite) antennas in a limited space \cite{pizzo2020spatially}. 

By using the channel degrees of freedom (DoFs), massive MIMO technology can significantly improve the channel capacity \cite{zhang20183d,yu2017theoretical}. With dense antenna deployment, holographic MIMO technology has the potential to further improve the channel capacity \cite{huang2020holographic}. To clarify the performance improvement brought by holographic MIMO, it is necessary to analyze the channel model and capacity of the holographic MIMO system.

In \cite{pizzo2020spatially}, a holographic MIMO small-scale fading model in Fourier plane-wave spectral representation is proposed. And then, the channel model in Fourier expansion form and the method of generating non-isotropic channels are given in \cite{pizzo2022fourier,pizzo2020holographic}. Based on the channel model proposed in \cite{pizzo2020spatially}, the authors of \cite{pizzo2020degrees} analyzed the upper limit of freedom of holographic MIMO in the wavenumber domain. In \cite{demir2022channel}, the spatial correlation matrices of holographic MIMO systems with different antenna spacing are compared. To analyze the performance of the continuous source and destination, the authors of \cite{wan2021capacity} derive the capacity bound for parallel linear source and destination based on electromagnetic information theory.

As discussed, existing works are mainly based on ideal assumptions. However, the angle distribution and array aperture constrain the capacity in reality. Therefore, this letter aims to analyze the realistic capacity of holographic MIMO channels with practical angle distribution and array aperture constraints. First, we use the transformation of coordinates in \cite{pizzo2020spatially} to calculate the spectral density in the wavenumber domain for generalized angle distribution. Then, mathematical expressions for the capacity of holographic MIMO channels are presented. Finally, simulation results show the effects of angle spread, signal-to-noise ratio (SNR), propagation scenarios, array aperture, and antenna spacing on channel capacity.

\textit{Notations}: Fonts a, $\mathbf{a}$, and $\mathbf{A}$ represent scalars, vectors, and matrices, respectively. 
$n \sim \mathcal{N}_{\mathbb{C}}\left(0, \sigma^{2}\right)$ stands for a circularly-symmetric complex-Gaussian random variable with variance $\sigma^{2}$.
$\hat{\mathbf{x}}$, $\hat{\mathbf{y}}$ and $\hat{\mathbf{z}}$ are three orthonormal vectors. 
$\mathbf{A}^{H}$ denotes Hermitian of $\mathbf{A}$.
$\mathbb{E}\{\cdot\}$ is the expectation operator.
$\lambda_{i}(\mathbf{A})$, $\operatorname{rank}(\mathbf{A})$, and $\operatorname{tr}\left(A\right)$ represent i-th sorted eigenvalue, rank, and trace of $\mathbf{A}$ ,respectively.
$\operatorname{diag}\left(\mathbf{a}\right)$ denotes the diagonal matrix with elements from $\mathbf{a}$.

\section{Holographic MIMO Channel Modeling}

\subsection{Channel Model in Wavenumber Domain}

Consider a holographic MIMO communication system, where both the receiver (RX) and the transmitter (TX) are z-oriented planar arrays with a large number of elements arranged below half wavelength. RX and TX span the rectangular regions of xy-dimensions $L_{R,x}$, $L_{R,y}$ and $L_{S,x}$, $L_{S,y}$, respectively. The system model can be given as
\begin{equation}\label{ideal_rectifier_resistance}
\mathbf{y}=\sqrt{\rho G_{t} G_{r}} \mathbf{H} \mathbf{x}+\mathbf{n},
\end{equation}
where $\mathbf{y} \in \mathbb {C}^{{N}_{R}}$ and $\mathbf{x} \in \mathbb {C}^{{N}_{S}}$ denote the received and transmitted signal vectors, respectively. $\mathbf{n} \in \mathbb {C}^{{N}_{R}}$ accounts for thermal noise that is distributed as $\mathbf{n} \sim \mathcal{N}_{\mathbb{C}}\left(\mathbf{0}, \sigma^{2} \mathbf{I}_{N_{R}}\right)$.
$\mathbf{H} \in \mathbb{C}^{N_{R} \times N_{S}}$is the small-scale fading channel matrix. $G_{t}$ and $G_{r}$ are the antenna gain of TX and RX, respectively, and $\rho$ is the SNR.

For an electromagnetic wave in arbitrary direction, its azimuth $\phi$ and elevation $\theta$ can be mapped to the wavenumber domain, i.e.,

\begin{equation}
    \left\{\begin{array}{l}
k_{x}=k \sin \theta \cos \phi \\
k_{y}=k \sin \theta \sin \phi \\
k_{z}=k \cos \theta
\end{array}\right.,
\end{equation}
where $k$ is the wavenumber, $k_{x}$, $k_{y}$ and $k_{z}$ denote the wavenumber in x, y and z directions, respectively.
Then the channel matrix $\mathbf{H}$ can be approximated as the superposition of the channel responses in the wavenumber domain \cite{pizzo2022fourier}, i.e.,

\begin{equation}\label{appromodel}
\mathbf{H} \approx \sqrt{N_{R}N_{S}}\mathbf{\Phi}_{r} \mathbf{H}_{a} \mathbf{\Phi}_{s}^{H},
\end{equation}
where

\begin{itemize}

\item
$\mathbf{\Phi}_{r}  \in \mathbb{C}^{N_{r} \times n_{R}}$ collects column vectors $\phi_{r}\left(l_{x}, l_{y}\right) \in \mathbb{C}^{N_{r} \times 1}$ with $n_{R}$ being the cardinality of the steering vector at RX. The $i$-th element of $\phi_{r}\left(l_{x}, l_{y}\right)$ is $\left[a_{r}\left(l_{x}, l_{y}, \mathbf{r}\right)\right]_{i}$, which can be expressed as 

\begin{center}
\begin{equation}
\begin{array}{l}
{\left[a_{r}\left(l_{x}, l_{y}, \mathbf{r}\right)\right]_{i}} \\
=\frac{1}{\sqrt{N_{r}}} e^{-\mathrm{j}\left(\frac{2 \pi}{L_{R,x}} l_{x} r_{x_{i}}+\frac{2 \pi}{L_{R,y}} l_{y} r_{y_{i}}+\gamma_{s}\left(l_{x}, l_{y}\right) r_{z_{i}}\right)}, \\
\end{array}    
\end{equation}
\end{center}

\item
$\mathbf{\Phi}_{s}\in\mathbb{C}^{N_{s} \times n_{S}}$ collects column vectors $\phi_{s}\left(m_{x}, m_{y}\right) \in \mathbb{C}^{N_{s} \times 1}$ with $n_{S}$ being the cardinality of the steering vector at TX. The $j$-th element of $\phi_{s}\left(m_{x}, m_{y}\right)$ is $\left[a_{s}\left(m_{x}, m_{y}, \mathbf{s}\right)\right]_{j}$, which can be expressed as 

\begin{center}
\begin{equation}
\begin{array}{l}
{\left[a_{s}\left(m_{x}, m_{y}, \mathbf{s}\right)\right]_{j}} \\
=\frac{1}{\sqrt{N_{s}}} e^{-\mathrm{j}\left(\frac{2 \pi}{L_{S,x}} m_{x} s_{x_{j}}+\frac{2 \pi}{L_{S,y}} m_{y} s_{y_{j}}+\gamma_{s}\left(m_{x}, m_{y}\right) s_{z_{j}}\right)}, \\

\end{array}    
\end{equation}
\end{center}

\item
$\mathbf{H}_{a} \in \mathbb{C}^{n_{R} \times n_{S}}$ is the angular random matrix collecting $H_{a}\left(\ell_{x}, \ell_{y}, m_{x}, m_{y}\right) \sim \mathcal{N}_{\mathbb{C}}\left(0, \sigma^{2}\left(\ell_{x}, \ell_{y}, m_{x}, m_{y}\right)\right)$, which is the angular response that maps source direction onto receive direction. $\sigma^{2}\left(\ell_{x}, \ell_{y}, m_{x}, m_{y}\right)$ represents the energy distribution of the channel in the wavenumber domain $\Omega_{R}\left(\ell_{x}, \ell_{y}\right)$ and $\Omega_{S}\left(m_{x}, m_{y}\right)$, i.e.,

\begin{equation}\label{nonideal_rectifier_resistance}
\begin{array}{l}
\left\{\left[\frac{2 \pi m_{x}}{L_{S, x}}, \frac{2 \pi\left(m_{x}+1\right)}{L_{S, x}}\right] \times\left[\frac{2 \pi m_{y}}{L_{S, y}}, \frac{2 \pi\left(m_{y}+1\right)}{L_{S, y}}\right]\right\}, \\
\left\{\left[\frac{2 \pi \ell_{x}}{L_{R, x}}, \frac{2 \pi\left(\ell_{x}+1\right)}{L_{R, x}}\right] \times\left[\frac{2 \pi \ell_{y}}{L_{R, y}}, \frac{2 \pi\left(\ell_{y}+1\right)}{L_{R, y}}\right]\right\}.
\end{array}
\end{equation}

\end{itemize}

Under the assumption that scattering is separable \cite{pizzo2022fourier}, the scattering decouples and $\sigma^{2}\left(\ell_{x}, \ell_{y}, m_{x}, m_{y}\right)$ becomes

\begin{equation}\label{nonideal_rectifier_resistance}
\sigma^{2}\left(\ell_{x}, \ell_{y}, m_{x}, m_{y}\right)=\sigma_{S}^{2}\left(m_{x}, m_{y}\right) \sigma_{R}^{2}\left(\ell_{x}, \ell_{y}\right).
\end{equation}
Then the angular random matrix $\mathbf{H}_{a} \in \mathbb{C}^{n_{R} \times n_{S}}$  can be obtained as \cite{pizzo2022fourier}

\begin{equation}\label{haapp}
\mathbf{H}_{a}=\operatorname{diag}\left(\boldsymbol{\sigma}_{R}\right) \mathbf{W} \operatorname{diag}\left(\boldsymbol{\sigma}_{S}\right),
\end{equation}
where $\boldsymbol{\sigma}_{R} \in \mathbb{R}_{+}^{n_{R}}$ and $\boldsymbol{\sigma}_{S} \in \mathbb{R}_{+}^{n_{S}}$ collect $\left\{\sigma_{R}\left(\ell_{x}, \ell_{y}\right)\right\}$ and $\left\{\sigma_{S}\left(m_{x}, m_{y}\right)\right\}$, respectively, and $\mathbf{W} \sim \mathcal{C N}\left(0, \mathbf{I}_{n_{r} n_{s}}\right)$.

\subsection {Numerical Solutions for Spectral Density in Wavenumber Domain}

The following is the calculation method about $\sigma_{R}\left(\ell_{x}, \ell_{y}\right)$, which also applies to $\sigma_{S}\left(m_{x}, m_{y}\right)$. We first let the polar coordinates $k_{R}=\frac{\sqrt{k_{x}^{2}+k_{y}^{2}}}{k}$ and $\phi_{R}=\phi$. The Jacobian of polar coordinates is given as

\begin{equation}
    \mathbf{J}\left(k_{R}, \phi_{R}\right)=\left[\begin{array}{ll}
\frac{\partial \theta}{\partial k_{R}} & \frac{\partial \theta}{\partial \phi_{R}} \\
\frac{\partial \phi}{\partial k_{R}} & \frac{\partial \phi}{\partial \phi_{R}}
\end{array}\right]=\left[\begin{array}{cc}
\frac{1}{\sqrt{1-k_{R}^{2}}} & 0 \\
0 & 1
\end{array}\right].
\end{equation}
Then $\sigma_{R}^{2}\left(\ell_{x}, \ell_{y}\right)$ can be calculated by the angle distribution function $f\left(\theta_{R}, \phi_{R}\right)$ as

\begin{equation}\label{bopujisuan}
\begin{split}
\sigma_{R}^{2}\left(l_{x}, l_{y}\right)&=\iint \limits_{\Omega_{R}\left(l_{x}, l_{y}\right)} f\left(\theta_{R}, \phi_{R}\right) d \theta_{R} d \phi_{R}\\&=\iint \limits_{\Omega_{R}\left(l_{x}, l_{y}\right)} f\left(\arcsin \left(k_{R}\right), \phi_{R}\right) \frac{1}{\sqrt{1-k_{R}^{2}}} d k_{R} d \phi_{R}.
\end{split}
\end{equation}

\begin{figure}[htbp]
  \begin{center}
  \includegraphics[width=3.0in]{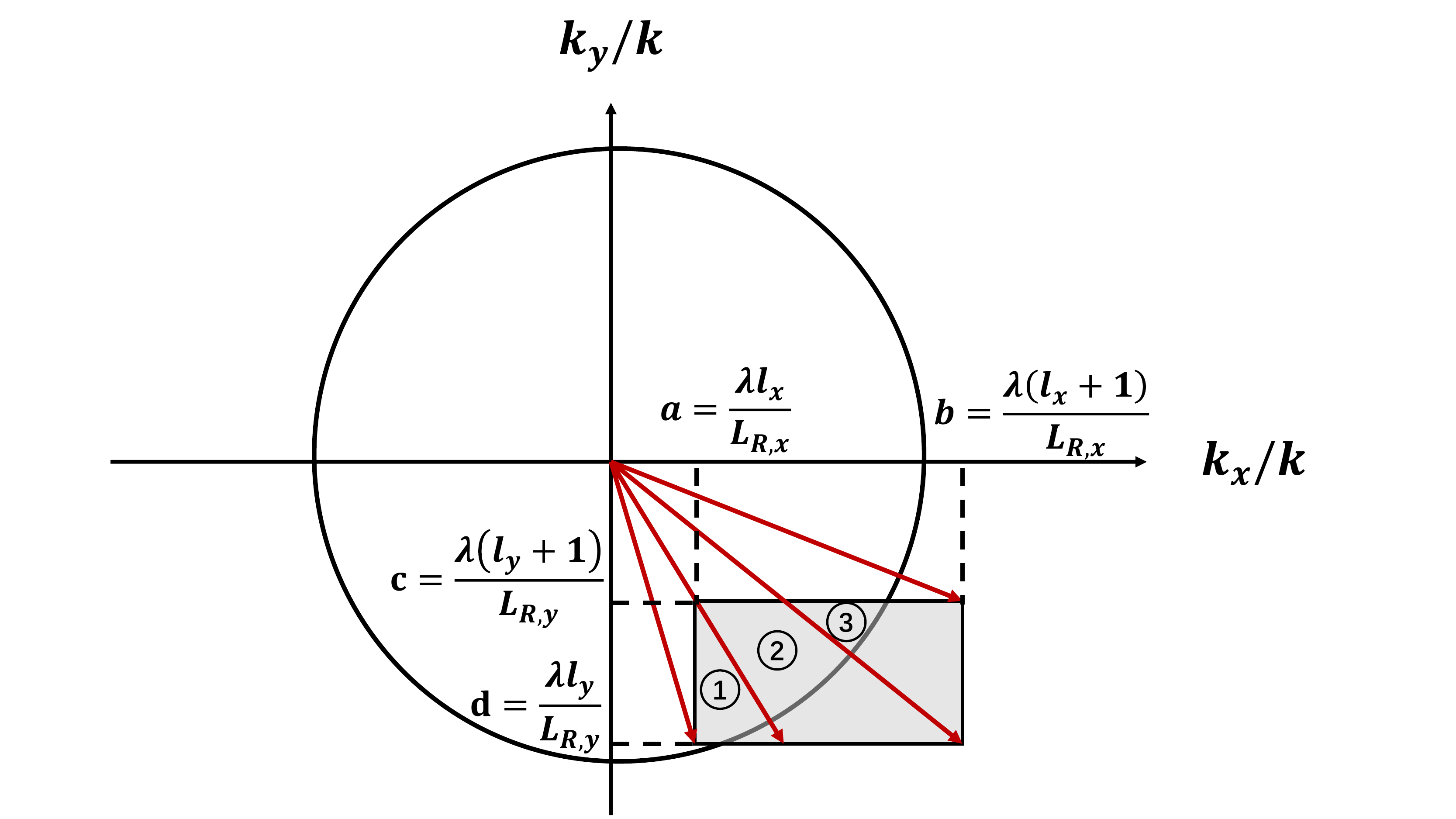}\\
  \caption{Three regions divided from the integral range of $\sigma_{R}^{2}\left(l_{x},l_{y}\right)$ in fourth quadrant when $|\ell_{x}| < |\ell_{y}|$.}\label{bopu}
  \end{center}
\end{figure}

The integral range $\Omega_{R}\left(\ell_{x}, \ell_{y}\right)$ can be divided into two or three regions under $\left(k_{R}, \phi_{R}\right)$ domain. The number of divided regions depends on whether the region is adjacent to the coordinate axis, and the numerical relationship between $|\ell_{x}|$ and $|\ell_{y}|$ will affect the division method \cite{pizzo2020spatially}. We take the fourth quadrant and $|\ell_{x}| < |\ell_{y}|$ for example as shown in Fig.\ref{bopu}, the integral region $\Omega_{R}\left(\ell_{x}, \ell_{y}\right)$ can be divided into three regions as \cite{pizzo2020spatially}
\begin{equation}\notag
    \begin{split}
\left\{(\phi_{R}, k_{R}) \mid 2\pi-\arctan{\frac{|c|}{a}} \leq \phi_{R} \leq 2\pi-\arctan{\frac{|d|}{a}}\right.,\\
\left.\min(1,\frac{a}{\cos{\left(2\pi-\phi\right)}}) \leq k_{R} \leq \min(1,\frac{|c|}{\sin{\left(2\pi-\phi\right)}})\right\},
    \end{split}
\end{equation}

\begin{equation}\notag
    \begin{split}
\left\{(\phi_{R}, k_{R}) \mid 2\pi-\arctan{\frac{|d|}{a}} \leq \phi_{R} \leq 2\pi-\arctan{\frac{|c|}{b}}\right.,\\
\left.\min(1,\frac{|d|}{\sin{\left(2\pi-\phi\right)}}) \leq k_{R} \leq \min(1,\frac{|c|}{\sin{\left(2\pi-\phi\right)}})\right\},
    \end{split}
\end{equation}

\begin{equation}\notag
    \begin{split}
\left\{(\phi_{R}, k_{R}) \mid 2\pi-\arctan{\frac{|c|}{b}} \leq \phi_{R} \leq 2\pi-\arctan{\frac{|d|}{b}}\right.,\\
\left.\min(1,\frac{|d|}{\sin{\left(2\pi-\phi\right)}}) \leq k_{R} \leq \min(1,\frac{b}{\cos{\left(2\pi-\phi\right)}})\right\}.
    \end{split}
\end{equation}

\begin{figure}[htbp]
  \begin{center}
  \includegraphics[width=3.0in]{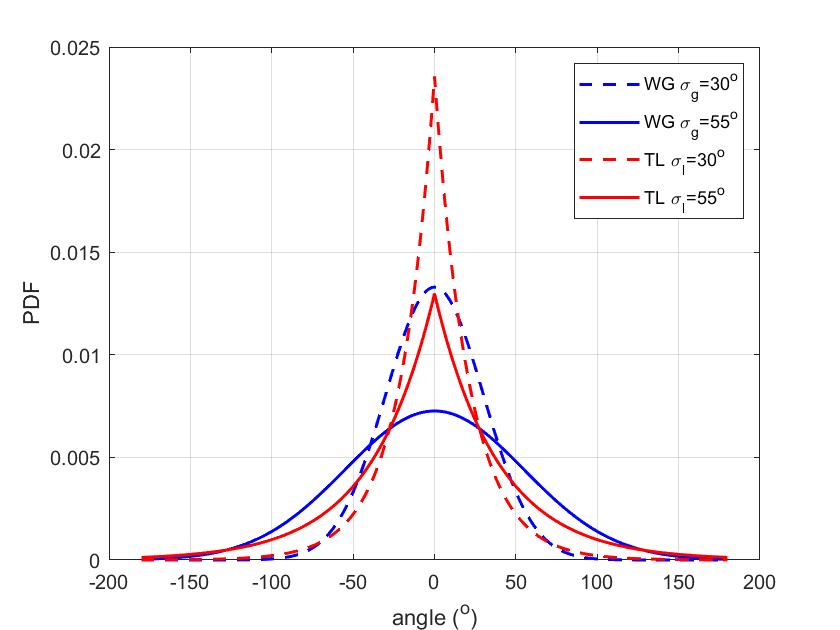}\\
  \caption{Probability density functions (PDFs) of WG and TL distributions.}\label{pdfwgtl}
  \end{center}
\end{figure}

Referring to the standard 3D MIMO channel model \cite{2022fp}, wrapped Gaussian (WG) and truncated Laplacian (TL) are used to model the azimuth angle distribution and the elevation angle distribution, respectively, and they are independent of each other. The corresponding curves are plotted in Fig.\ref{pdfwgtl}. Therefore, the angle distribution function $f\left(\theta, \phi\right)$ can be expressed in the following form.

\begin{equation}\label{diode_current_waveform_time_domain}
f(\theta, \phi)=\frac{Q_{l}}{\sqrt{2} \sigma_{l}} e^{-\frac{\sqrt{2}\lvert \theta-\theta_{0}\rvert }{\sigma_{l}}} \cdot \frac{Q_{g}}{\sqrt{2 \pi} \sigma_{g}} e^{-\frac{\left(\phi-\phi_{0}\right)^{2}}{2 \sigma_{g}^{2}}},
\end{equation}
where $Q_{l}$ and $Q_{g}$ are the normalization factors, $\phi_{0}$ and $\theta_{0}$ are central angle of the azimuth and elevation, $\sigma_{g}$ and $\sigma_{l}$ are variance of the azimuth and elevation. We only consider the wave propagation towards $-z$ axis at TX and RX, so the elevation $\theta\in[0, \pi / 2]$ and the azimuth $\phi\in[\phi_{0}-\pi, \phi_{0}+\pi]$.

\section{Performance Analysis}

We now use the channel model in Section II to analyze the capacity of the holographic MIMO system. Since $\mathbf{\Phi}_{r}$ and $\mathbf{\Phi}_{s}$ in \eqref{appromodel} are semi-unitary matrices\cite{pizzo2022fourier}, the holographic MIMO communication systems can be equivalently expressed as

\begin{equation}
\mathbf{y}_{a}=\sqrt{\rho G_{t} G_{r} N_{S} N_{R}} \mathbf{H}_{a} \mathbf{x}_{a}+\mathbf{n}_{a},
\end{equation}
where $\mathbf{y}_{a}=\boldsymbol{\Phi}_{r}^{\mathrm{H}} \mathbf{y} \in \mathbb{C}^{n_{R}}$ and $\mathbf{x}_{a}=\boldsymbol{\Phi}_{s}^{\mathrm{H}} \mathbf{x} \in \mathbb{C}^{n_{S}}$ denote the received and transmitted signal vectors in the wavenumber domain, respectively. $\mathbf{n}_{a}=\boldsymbol{\Phi}_{r}^{\mathrm{H}} \mathbf{n} \in \mathbb{C}^{n_{R}}$
is the angular noise vector distributed as $\mathbf{n}_{a} \sim \mathcal{N}_{\mathbb{C}}\left(\mathbf{0}, \mathbf{I}_{n_{R}}\right)$.

\subsection{Capacity under Generalized Angle Distributions}

In the case of unknown channel state information at the transmitter, the capacity can be obtained as \cite{pizzo2022fourier}

\begin{equation}\label{cap}
C=\sum_{i=1}^{\operatorname{rank}(\mathbf{H}_{a})} \mathbb{E}\left\{\log _{2}\left(1+\frac{\rho G_{t} G_{r} N_{S} N_{R} }{n_{S}} \times \lambda_{i}(\mathbf{H}_{a}\mathbf{H}_{a}^{\mathrm{H}})\right)\right\},   
\end{equation}
where $\lambda_{i}(\mathbf{H}_{a}\mathbf{H}_{a}^{\mathrm{H}})$ is $i$-th sorted eigenvalue of $\mathbf{H}_{a}\mathbf{H}_{a}^{\mathrm{H}}$, and $n_{S}\approx \frac{\pi {L}_{S,x} {L}_{S,y}}{\lambda^{2}}$ denotes the cardinality of the steering vector at TX \cite{pizzo2020degrees}.

By substituting \eqref{haapp} into \eqref{cap}, the channel capacity with generalized angle distributions can be expressed as 

\begin{equation}
\begin{split}    
C=&\sum_{i=1}^{\operatorname{rank}(\mathbf{A})} \mathbb{E}\Big\{\log _{2}\Big(1+\frac{\rho G_{t} G_{r} N_{S} N_{R} }{n_{S}} \times \Big. \Big. \\
&\Big. \Big. \lambda_{i}(\operatorname{diag}\left(\boldsymbol{\sigma}_{R} \odot \boldsymbol{\sigma}_{R}\right) \mathbf{W} \mathbf{W}^{\mathrm{H}} \operatorname{diag}\left(\boldsymbol{\sigma}_{S} \odot \boldsymbol{\sigma}_{S}\right))\Big)\Big\},
\end{split}
\end{equation}
where the elements of $\boldsymbol{\sigma}_{R}$ and $\boldsymbol{\sigma}_{S}$ can be obtained by \eqref{bopujisuan}.

\subsection{Capacity for Special Cases}

It can be seen from \eqref{cap} that the channel capacity depends on the eigenvalue distribution of $\mathbf{H}_{a}\mathbf{H}_{a}^{\mathrm{H}}$ when the antenna configuration is fixed. According to Jensen's inequality \cite{tse2005fundamentals}, the channel capacity reaches the upper bound when the eigenvalues are uniformly distributed, which corresponds to isotropic scattering. Assuming that the total power gain is 1, i.e., $\operatorname{tr}\left(\mathbf{H}_{a} \mathbf{H}_{a}^{\mathrm{H}}\right)=1$, the upper bound is 

\begin{equation}
    C_{upper}=n_{M}\log _{2}\left(1+\frac{\rho G_{t} G_{r} N_{S} N_{R}}{n_{S} n_{M}}\right),
\end{equation}
where $n_{M}$ is $\operatorname{min}\left(n_{S},n_{R}\right)$.

It is difficult to satisfy isotropic propagation in reality, so there is a gap between the upper bound and the channel capacity under generalized angle distributions. However, there is a special case for the gap. According to the equivalent infinitesimal $\ln{\left (1+x  \right )} \sim x$, when SNR is very low, i.e., $\rho \longrightarrow 0$, the capacity can be observed as
\begin{equation}\label{dengjiawuqiong}
\begin{aligned}
C_{low-snr}&=\sum_{i=1}^{\operatorname{rank}\left(\mathbf{H}_{a}\right)} \mathbb{E}\left\{\log _{2}\left(1+\frac{\rho G_{t} G_{r} N_{R} N_{S}}{n_{S}}  \lambda_{i}\left(\mathbf{H}_{a} \mathbf{H}_{a}^{\mathrm{H}}\right)\right)\right\} \\
&\approx\sum_{i=1}^{\operatorname{rank}\left(\mathbf{H}_{a}\right)} \mathbb{E}\left\{\frac{\rho G_{t} G_{r} N_{R} N_{S}}{n_{S} \ln{2}}  \lambda_{i}\left(\mathbf{H}_{a} \mathbf{H}_{a}^{\mathrm{H}}\right)\right\} \\
&=\frac{ \rho G_{t} G_{r} N_{R} N_{S} }{n_{S} \ln{2}} \sum_{i=1}^{\operatorname{rank}\left(\mathbf{H}_{a}\right)} \mathbb{E}\left\{\lambda_{i}\left(\mathbf{H}_{a} \mathbf{H}_{a}^{\mathrm{H}}\right)\right\} \\
&= \frac{\rho G_{t} G_{r} N_{R} N_{S} }{n_{S} \ln{2}} \mathbb{E}\left\{\operatorname{tr}\left(\mathbf{H}_{a} \mathbf{H}_{a}^{\mathrm{H}}\right)\right\} \\
&= \frac{\rho G_{t} G_{r} N_{R} N_{S} }{n_{S} \ln{2}}.
\end{aligned}
\end{equation}
Thus the channel capacity will no longer depend on angle distributions and the gap will approximately disappear when SNR tends to zero.

\begin{figure}[htbp]
  \begin{center}
  \includegraphics[width=3.0in]{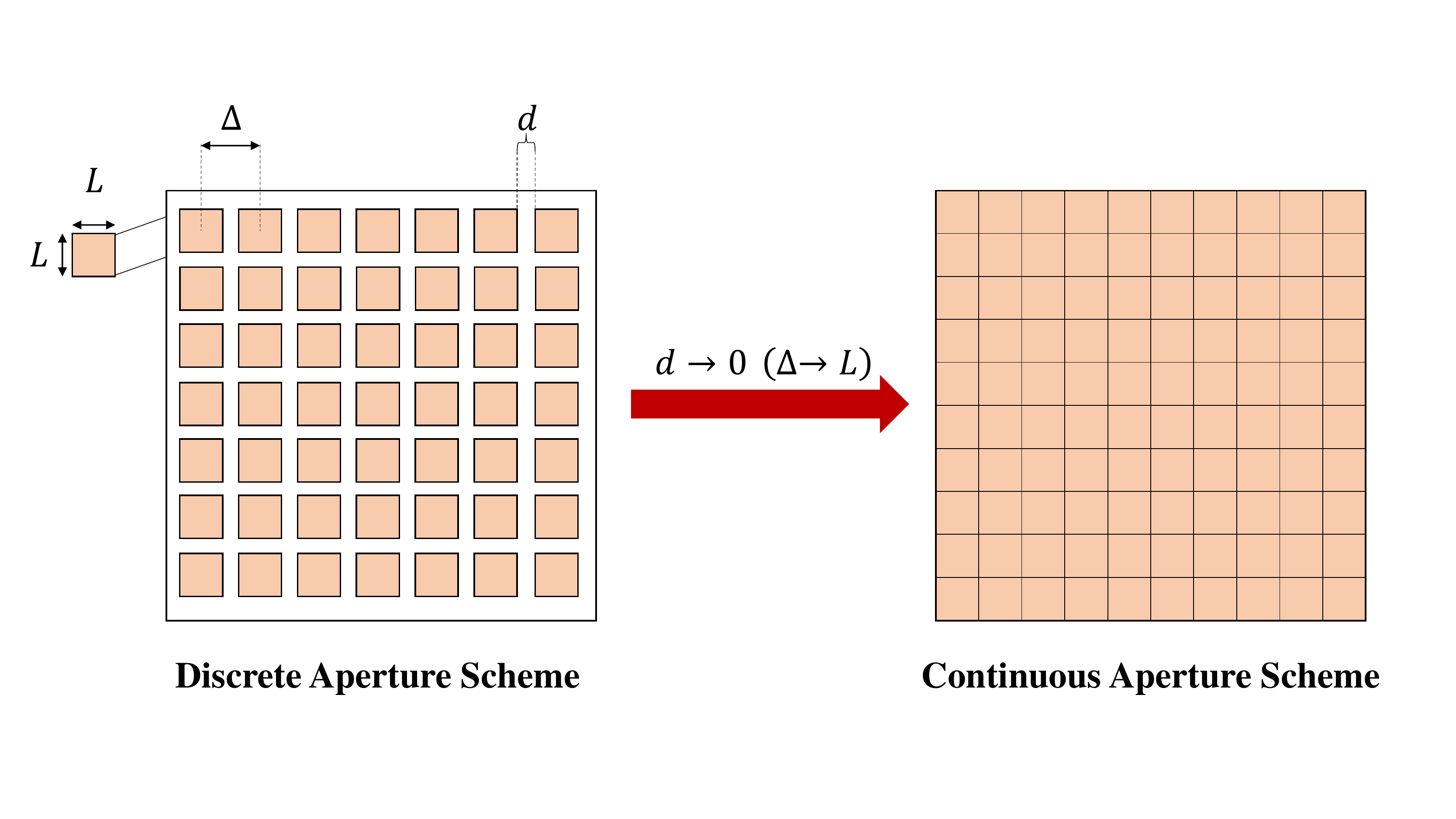}\\
  \caption{Discrete and continuous aperture schemes.}\label{scheme}
  \end{center}
\end{figure}

\subsection{Capacity for Different Aperture Schemes}

A convenient implementation of holographic MIMO is discrete aperture scheme, in which the horizontal or vertical antenna spacing is usually below half of the wavelength. However, the spatial utilization of discrete aperture scheme is limited. To improve the performance with limited apertures, continuous aperture scheme is proposed, in which the antenna spacing $d$ tends to zero \cite{2022Extremely}. Fig.~\ref{scheme} illustrates the characteristics and relationships of the two different aperture schemes.

In the discrete aperture scheme, the antenna gain of each patch antenna is a constant and can be calculated by

\begin{equation}\label{antennagain}
    G=\frac{4 \pi \eta S}{\lambda^{2}},
\end{equation}
where $\eta \textless$1 is aperture efficiency, and $S$ is antenna area, and $\lambda$ is the operating wavelength \cite{baranis1982antenna}. Then the capacity of the holographic MIMO system in the discrete aperture scheme can be given by 

\begin{equation}
\begin{split}    
C=&\sum_{i=1}^{\operatorname{rank}(\mathbf{A})} \mathbb{E}\Big\{\log _{2}\Big(1+\frac{16 \pi^{2} \rho \eta_{t} \eta_{s} A_{R} A_{S} N_{R} N_{S}}{n_{S} \lambda^{4}} \times \Big. \Big. \\
&\Big. \Big. \lambda_{i}(\operatorname{diag}\left(\boldsymbol{\sigma}_{R} \odot \boldsymbol{\sigma}_{R}\right) \mathbf{W} \mathbf{W}^{\mathrm{H}} \operatorname{diag}\left(\boldsymbol{\sigma}_{S} \odot \boldsymbol{\sigma}_{S}\right))\Big)\Big\},
\end{split}
\end{equation}
where $\eta_{t}$ and $\eta_{s}$ denote the aperture efficiency at TX and RX, $A_{R}$ and $A_{S}$ denote the antenna area at TX and RX, respectively.

In the continuous aperture scheme, the area of each antenna at TX and RX is  $A_{S}=\frac{L_{S, x} L_{S, y}}{N_{S}}$ and $A_{R}=\frac{L_{R, x} L_{R, y}}{N_{R}}$, respectively. Therefore, the capacity of the holographic MIMO system in the continuous aperture scheme can be given by

\begin{equation}
\begin{split}    
C=&\sum_{i=1}^{\operatorname{rank}(\mathbf{A})} \mathbb{E}\Big\{\log _{2}\Big(1+\frac{16 \pi^{2} \rho \eta_{t} \eta_{s} L_{S, x} L_{S, y} L_{r, x} L_{r, y}}{n_{S} \lambda^{4}} \times \Big. \Big. \\
&\Big. \Big. \lambda_{i}(\operatorname{diag}\left(\boldsymbol{\sigma}_{R} \odot \boldsymbol{\sigma}_{R}\right) \mathbf{W} \mathbf{W}^{\mathrm{H}} \operatorname{diag}\left(\boldsymbol{\sigma}_{S} \odot \boldsymbol{\sigma}_{S}\right))\Big)\Big\}.
\end{split}
\end{equation}

It can be seen that the capacity of continuous aperture scheme depends on the total area of TX and RX, and the capacity will not increase with more antennas in a limited space due to the constraint between the area and number of antenna.

\section{Numerical Results}

In this section, we numerically analyze the performance of holographic MIMO system based on the theoretical analysis in section III. According to the channel model standards 3GPP TR38.901 \cite{3gpp_38.901}, three scenarios are selected to analyze the influence of different scenarios on the performance of the holographic MIMO system. Table \ref{tab:my_label} shows the default simulation parameters.

\begin{table}[htbp]
    \caption{Simulation parameters}
    \label{tab:my_label}
    \centering
    \begin{tabular}{l l l l }
         \toprule[1pt]
         Operating frequency & \multicolumn{3}{c}{6 GHz}\\
         \midrule[0.5pt]
         Mean azimuth & \multicolumn{3}{c}{90$^{\circ}$}\\
         \midrule[0.5pt]
         Mean elevation & \multicolumn{3}{c}{45$^{\circ}$}\\
         \midrule[0.5pt]
         Antenna size & \multicolumn{3}{c}{$\lambda/$8 $\times$ $\lambda/$8}\\
         \midrule[0.5pt]
         Aperture efficiency & \multicolumn{3}{c}{60$\%$}\\
         \midrule[0.5pt]
         Scenarios (LOS) & UMa & UMi &RMa\\
         \midrule[0.5pt]
         Azimuth angle spread at transmitter $\sigma_{sg}$ & 14.0$^{\circ}$ & 14.7$^{\circ}$ & 7.9$^{\circ}$\\
         \midrule[0.5pt]
         Elevation angle spread at transmitter $\sigma_{sl}$ & 0.3$^{\circ}$ & 0.6$^{\circ}$ & 0.1$^{\circ}$\\
         \midrule[0.5pt]
         Azimuth angle spread at receiver $\sigma_{rg}$ & 65$^{\circ}$ & 46$^{\circ}$ & 33$^{\circ}$\\
         \midrule[0.5pt]
         Elevation angle spread at receiver $\sigma_{rl}$ & 8.9$^{\circ}$ & 4.4$^{\circ}$ & 3.0$^{\circ}$\\
         \bottomrule[1pt]
    \end{tabular}
\end{table}

\begin{figure}[htbp]
  \begin{center}
  \includegraphics[width=3.0in]{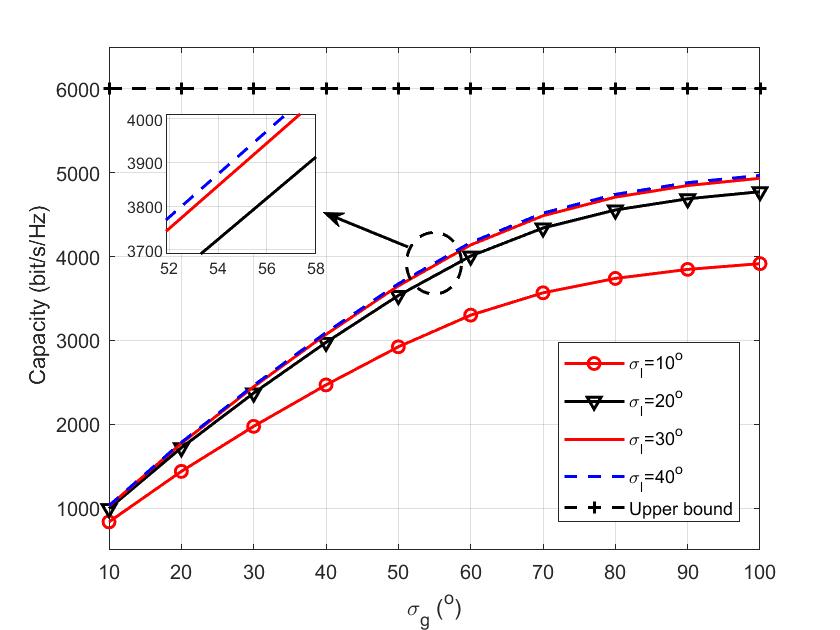}\\
  \caption{Capacity in bit/s/Hz as a function of azimuth angle spread in degree under different elevation angle spread.}\label{as}
  \end{center}
\end{figure}

Fig.~\ref{as} depicts the performance of holographic MIMO versus angle spread. The array aperture for both TX and RX is set as  15 $\lambda$$\times$15 $\lambda$. The SNR is 30 dB and the antenna spacing is $\lambda/$4. We can find in Fig.~\ref{as} that when the elevation angle spread is fixed, the capacity will increase with the azimuth angle spread. When the azimuth angle spread is large enough, e.g., 80$^{o}$, the capacity will increase slowly and still have a gap with the upper bound. Also, the capacity will increase fast with elevation angle spread when the elevation angle spread is small but increase slowly when the elevation angle spread is large. This is due to the mapping relationship between angle distribution and wavenumber distribution. 



\begin{figure}[htbp]
  \begin{center}
  \includegraphics[width=3.0in]{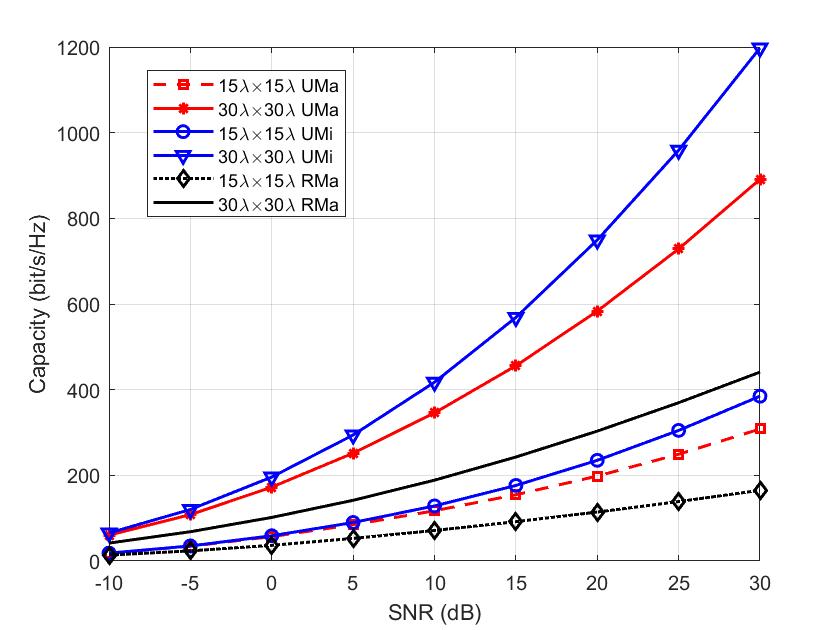}\\
  \caption{Capacity in bit/s/Hz as a function of SNR in dB under different scenarios and array aperture.}\label{angleapertureplot}
  \end{center}
\end{figure}

Fig.~\ref{angleapertureplot} compares the performance of holographic MIMO under different scenarios and array aperture. Both antenna spacing of TX and RX is $\lambda/$4. The array aperture is set to two different values, i.e., 15 $\lambda$$\times$15 $\lambda$ and 30 $\lambda$$\times$30 $\lambda$. It can be seen that the holographic MIMO system under UMi scenario obtains the largest channel capacity and under RMa scenario obtains the smallest channel capacity. This is because UMa and UMi scenarios have more diffuse angle distribution than RMa scenario. And the capacity gap is not obvious when the SNR is low, which consists with \eqref{dengjiawuqiong}. Besides, the holographic MIMO system with 30 $\lambda$$\times$30 $\lambda$ array obtains more capacity than 15 $\lambda$$\times$15 $\lambda$ array under the same scenario. This is because the larger aperture brings greater angular resolution, which improves the multiplexing gain and hence improves the capacity.

\begin{figure}[htbp]
  \begin{center}
  \includegraphics[width=3.0in]{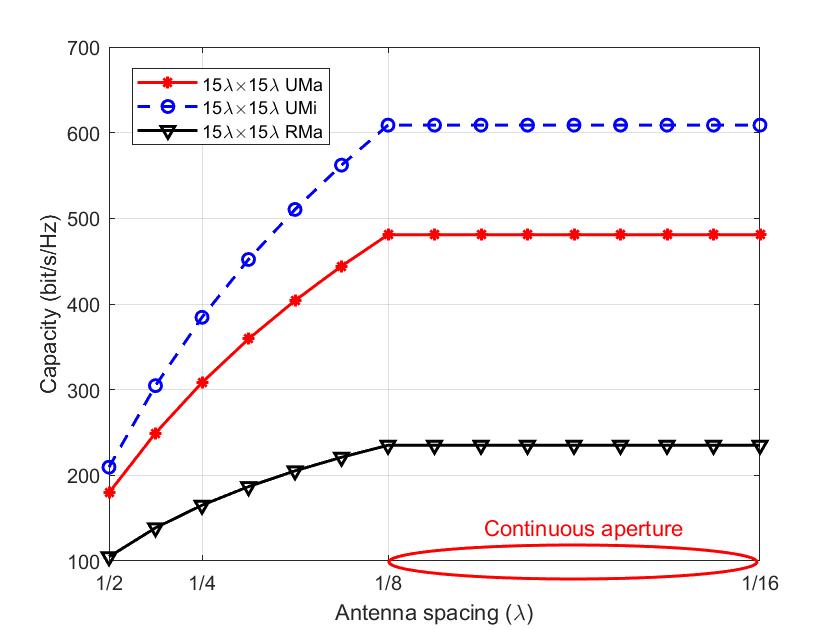}\\
  \caption{Capacity in bit/s/Hz as a function of normalized antenna spacing under different scenarios.}\label{antennaspacingplot}
  \end{center}
\end{figure}

Fig.~\ref{antennaspacingplot} depicts the effect of antenna spacing on channel capacity. We compare the capacity curves under different propagation scenarios. The array aperture for both TX and RX is set as 15 $\lambda$$\times$15 $\lambda$. As illustrated in Fig.~\ref{antennaspacingplot}, when antenna spacing $\Delta$ is not smaller than $\lambda/$8, the channel capacity increases with the decrease of antenna spacing. This is because decreased antenna spacing results in a larger transmitting and receiving surface. However, when antenna spacing $\Delta$ is $\lambda/$8, the number of deployed antennas reaches an upper limit, and the aperture tends to be continuous. If further decreasing the antenna spacing, the area of each antenna should be reduced. In this case, the capacity is no longer increasing because the effect of antenna gain cancels out with the effect of antenna density. It should be noted that the aperture efficiency is assumed to be constant. In practice, aperture efficiency may change with the decrease in antenna size and therefore affects the capacity value, which needs to be verified by measurement.

\section{Conclusion}
This letter analyzed the capacity of holographic MIMO channels with practical constraints. We focused on the capacity constrained by generalized angle distribution and limited array aperture. Theoretical analysis and numerical results show that angle distribution and array aperture are the key factors affecting channel capacity. Due to the generalized angle distribution, the realistic capacity with sufficient angle spread cannot reach the upper bound. In addition, the angle distribution affects the capacity at high SNR more obviously than the capacity at low SNR. Besides, smaller antenna spacing with a limited aperture improves the capacity in the discrete aperture scheme but not improves the capacity in the continuous aperture scheme.


%

\bibliographystyle{IEEEtran}
\bibliography{IEEEabrv,Bibliography}


\newpage

\vfill

\end{document}